\documentclass[epj]{svjour}
\usepackage{graphics}
\usepackage{graphicx}
\usepackage{dcolumn}
\usepackage{bm}
\usepackage{color}

\newcommand{\be}{\begin{equation}}
\newcommand{\ee}{\end{equation}}
\newcommand{\ba}{\begin{eqnarray}}
\newcommand{\ea}{\end{eqnarray}}
\newcommand{\bd}{\begin{displaymath}}
\newcommand{\ed}{\end{displaymath}}
\newcommand{\bea}{\begin{eqnarray}}
\newcommand{\eea}{\end{eqnarray}}

\renewcommand{\vec}[1]{\mbox{\boldmath$#1$}}

\begin{document}
\title{Investigate the $\Lambda$ and $\bar{\Lambda}$ polarization splitting effect with combined mechanisms}

\author{Simin Wu\inst{1} \and Yilong Xie\inst{1}
}


\institute{School of Mathematics and Physics, China University of Geosciences (Wuhan), Lumo Road 388, 430074 Wuhan, China}
\date{Received: date / Revised version: date}
\abstract{The significant splitting of $\Lambda$ and $\bar{\Lambda}$ polarization measured in STAR's Au+Au 7.7GeV collisions seems to be huge and unable to be described satisfactorily by any single mechanism, thus we revisit and combine there different mechanisms together on the basis of our PICR hydrodynamic model, to explain the experimental data. The three mechanisms, i.e. the meson field mechanism, the freeze-out space-time mechanism, and the QGP's magnetic field mechanism, lie on different stage of high energy collisions, and thus are not contradicted with each other. We find that the meson field mechanism is dominat, while the QGP's magnetic field mechanism is rather trivial, and freeze-out time effect is restricted by the small FZ time difference, leading to a hierarchy of $\Delta P_J \gg \Delta P_t \gg \Delta P_m$. Besides, the combination of different mechanisms could promote the mean value of polarization splitting from about 3\%-4\% to 4.5\%, which is more close to the experimental measured mean value of 5.8\%.
}
\maketitle
\section{Introduction}\label{Intro}

Experiments at RHIC and many other facilities have evidently revealed that the hot-dense matter created in non-central heavy ion collisions, i.e. quark gluon plasma (QGP), is rotating extremely fast with the vorticity of order $10^{21} s^{-1}$ \cite{Nature,SecondSTAR,ALICE,HADES}.
The rotating QGP, with a huge initial angular momentum of order $10^5\hbar$ \cite{BPR2008,GCD2008,VAC2014} lead to various phenomena, such as Chiral Vortical Effect(CVE) and hyperon spin polarization \cite{GLP2012}. The spin polarization of $\Lambda$ hyperon detected by RHIC has triggered tremendous discussions recently. The underlying mechanism is similar to the the Einstein-de-Hass effect \cite{EH1915} and Barnet effect \cite{B1935}: the particles' spin will be aligned to the initial angular momentum or vorticity, under spin-orbital coupling effect \cite{BCDG13,BCW2013}.

The STAR Collaboration at RHIC has measured $\Lambda$ and $\bar{\Lambda}$ polarization for Au+Au collisions at different energies $\sqrt{S_{NN}}=7.7-200$ GeV. While the results confirmed many earlier theoretical predictions by Becattini and Wang et al. \cite{BCDG13,BCW2013,LW05a,HHW11,BGT07}, the difference between the polarization of $\bar{\Lambda}$ and $\Lambda$ was not expected. Specifically the magnitude of $\bar{\Lambda}$ polarization is always greater than that of $\Lambda$ polarization.
This splitting effect has raised great interests and then various mechanisms were proposed to explained \cite{CKW2019,XCC2021,VBZ2020,GLW2019,I2020,ATC2020,XLH2022,HX2018,BGS2018,FPW2016}.
E.g.,1) the ref. \cite{CKW2019} proposed the mechanism of meson field, which is caused by the baryon vorticity, and our recent work \cite{XCC2021} has also dug into this mechanism, 2) the ref.\cite{VBZ2020} proposed that the freeze-out space-time of $\Lambda$ and $\bar{\Lambda}$ plays role in the splitting phenomenon, 3) the ref. \cite{GLW2019} proposed the magnetic field caused by the charged QGP vorticity also contribute to the splitting. The above 3 mechanisms lie on the different stage of collision system, meaning that they are not contradicted with each other. Actually they are correlated to some extents, e.g. if the the freeze-out time of $\Lambda$ and $\bar{\Lambda}$ are different, then the meson field and QGP's magnetic field at freeze-out will be different for $\Lambda$ and $\bar{\Lambda}$, leading to an enhancement of the polarization difference. Since the freeze-out time difference will be small, the estimated correlations between the above 3 mechanism will not be significant, and thus we will not consider the correlation effects in this work.

Due to the large errors in experimental measurements, the polarization splitting can not be clearly recognized, but at least for collision energy of $\sqrt{s_{\rm NN}}=$ 7.7 GeV, the splitting effect can be identified with high confidence level. Actually, the experimentally measured splitting at 7.7 GeV gives an average value of about 6\% (with 3.5\% statistical uncertainties), which can not be approached by any single mechanism by far 
and hints the call for combined mechanism to explain. On the other hand, all these mechanisms are investigated by using different phenomenological models, including the above mentioned 3 mechanisms. Therefore, in this work we are going to revisit the above 3 mechanisms, and then combine them together, by using a uniform model, i.e. our high resolution (3+1)D particle-in-cell relativistic (PICR) hydrodynamic model, to investigate the splitting effect of $\Lambda$ and $\bar{\Lambda}$ polarization.

The paper is organized as follows. In Section \ref{S2}, we revisit the magnetic field of QGP and the meson magnetic field, and Section \ref{S3} is dedicated to the freeze-out time mechanism. In Section \ref{S4}, we combine the 3 mechanisms together, as well as some disccusions. Summary and conclusions are drawn in Section \ref{S5}.

\section{Magnetic field of charged QGP and vector meson in a rotating system}
\label{S2}

\begin{figure}[ht] 
\begin{center}
      \includegraphics[width=4cm]{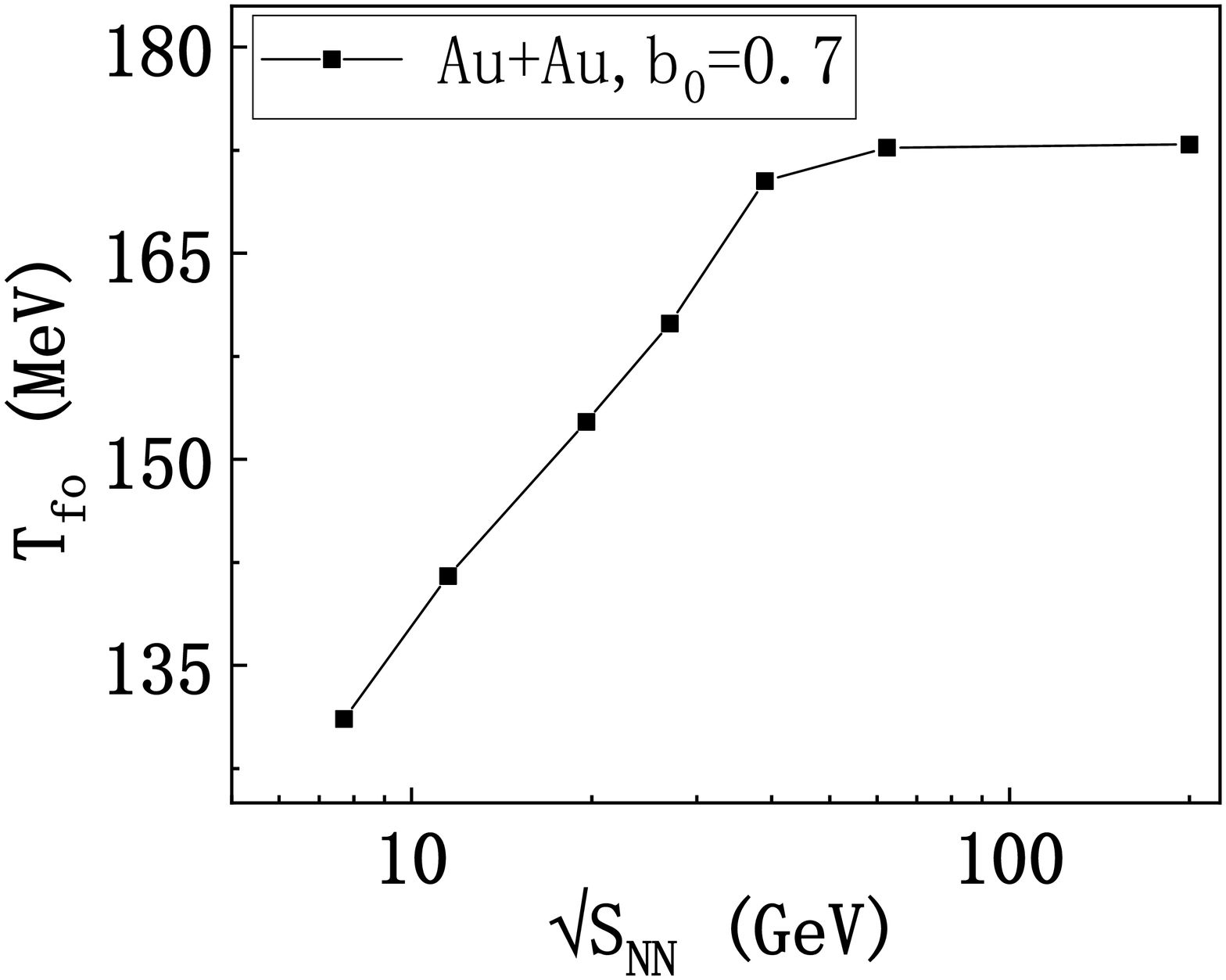}
      \includegraphics[width=4cm]{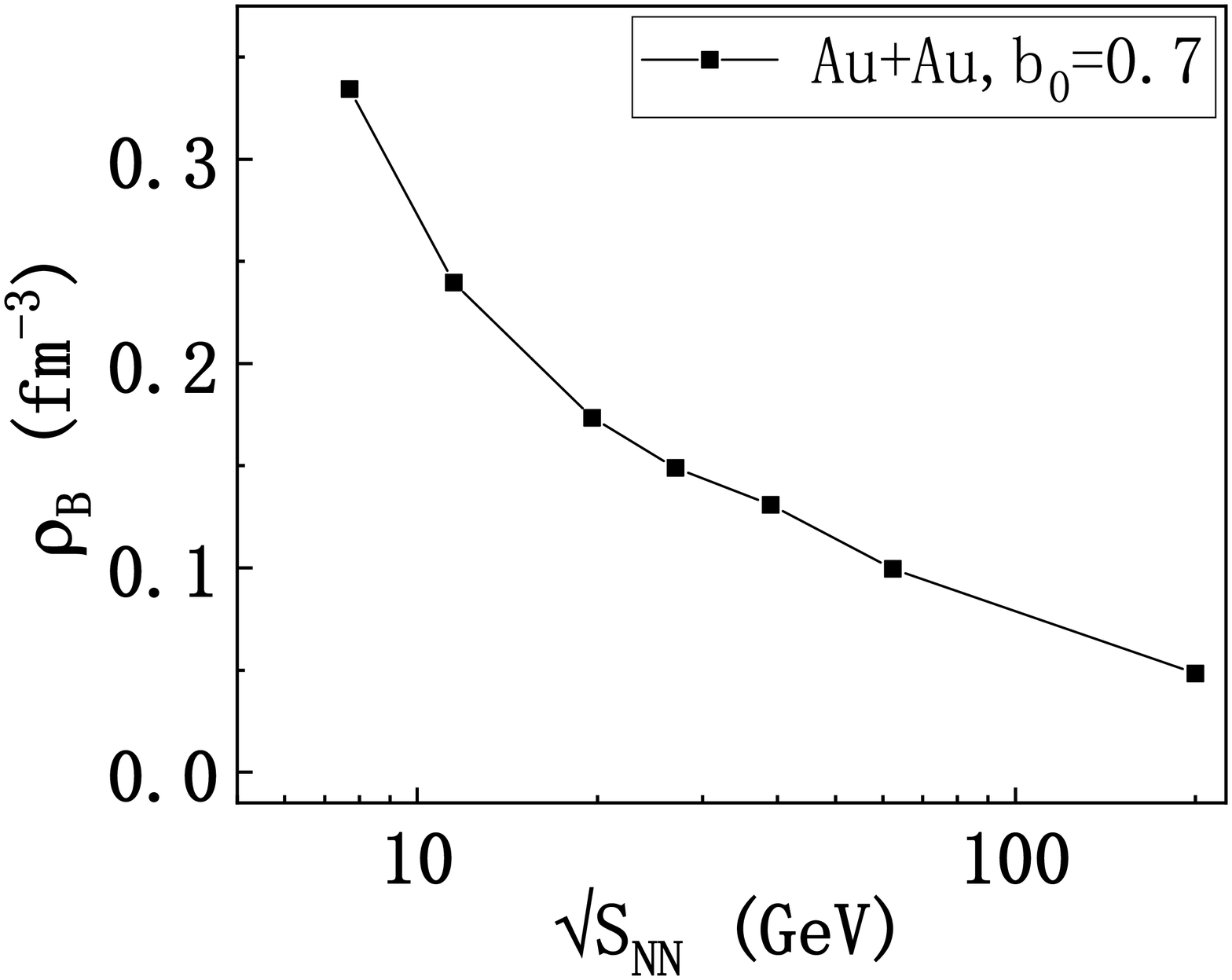}
\end{center}
\vspace{-0.3cm}
\caption{
(Color online)
The average freeze-out temperature and baryon density at different collision energy $\sqrt{S_{NN}}=7.7-200$ GeV, with fixed $b_0=0.7$.
}
\label{TempDen}
\end{figure}

As mentioned in the Introduction, the magnetic field can be generated naturally in a charged vortical QGP. On the other hand the magnetic field  of vector meson is also expected to be built by the vortical baryon current. In general, if the fermions are rotating along with a fluid system, the bosons, e.g. massless photon or massive vector meson, that mediate the interaction between fermions, will have a nonvanishing magnetic component of bosonic field. And as is known, the magnetic field always offers opposite aligned directions to the spin of particles and anti-particles, thus leading to the polarization splitting. In this Section, the above two kinds of magnetic field and their contributions to the splitting are investigated by our PICR hydrodynamic model, which has been successfully used in our previous polarization studies \cite{XCC2021,Xie2016,Xie2017,XWC2019}. 

In this work, the collisions at different energies $\sqrt{S_{NN}}=7.7-200$ GeV are simulated by the PICR hydrodynamic model, and we use the same impact parameter ratio as Ref. \cite{XCC2021}: $b_0=b/b_{max}=0.45-0.7$ (where b is the impact parameter and the $b_{max}$ is the maximum impact parameter), the centrality is estimated as $c\approx b_0^2$. The time increment is $\Delta t \approx 0.0423 fm/c$; the cell size is $0.342^3 fm^3$. The freeze-out time increases from $5.9fm/c$ to $7.9fm/c$ with increasing collision energy 7.7-200GeV, so that the average freeze-out temperature and the baryon density of the system in our model, as shown in Fig. \ref{TempDen}, agrees with the experimental results and theoretical models\cite{BMG2006,STAR2013}. Besides, the electric charge-to-baryon ratio is fixed $n_Q=0.4n_B$ as in the Ref. \cite{MSS2019}.

\begin{figure}[ht]
\begin{center}
\includegraphics[width=8.5cm]{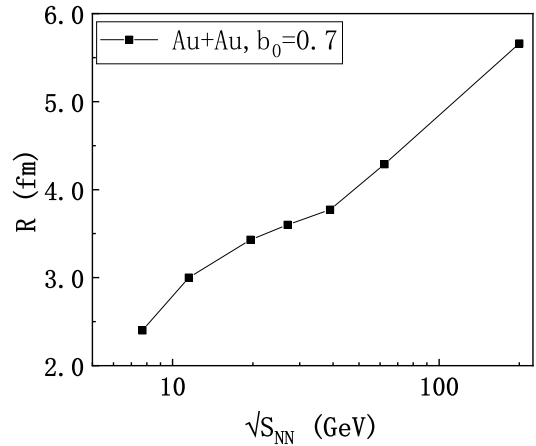}
\end{center}
\caption{
(Color online)
The average radius on reaction plane($y = 0$) for different collision energy at freeze-out in our PICR hydrodynamics model.
}
\label{radius}
\end{figure}

\subsection{Magnetic field}\label{S2.1}

According to Ref. \cite{GLW2019}, the in-medium long-life magnetic field arises along with the rotated QGP system can contribute to the difference between $\Lambda$ and $\bar{\Lambda}$ polarization. The collision system therein was considered as a many-body fluid system that has nonzero vorticity and charge density, but the system shape was assumed to be a simple planar swirl, just like a disk (seen in Fig.1 of Ref.\cite{GLW2019}). Therefore in this part of our work, a spherical, instead of planar, fluid system that is uniformly charged,is assumed. Although, the true system will be more like an ellipsoid, we hope the assumption of sphere will be close enough to the physical reality. To determine the radius of the sphere, we average the spans of $x$ and $z$ directions on the reaction plane (at freeze-out), the result is shown in Fig. \ref{radius}.

\begin{figure}[ht]
\begin{center}
\includegraphics[width=0.5\textwidth]{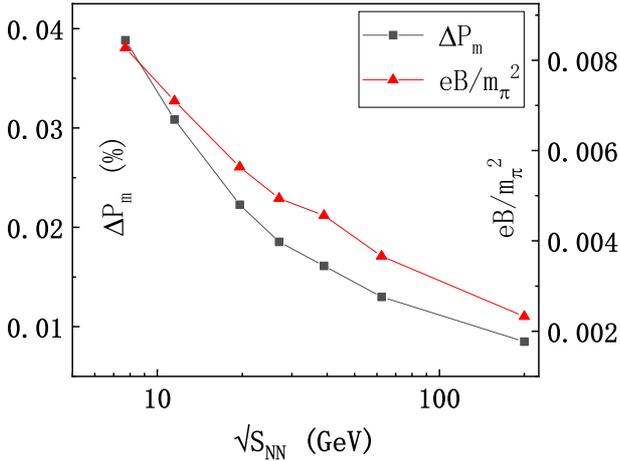}
\end{center}
\caption{
(Color online)
The polarization splitting $\Delta P_m$ between $\Lambda$ and $\bar{\Lambda}$ induced by QGP's magnetic field, for different Au+Au collisions energies $\sqrt{S_{NN}}=7.7-200$ GeV with fixed impact parameter ratio $b_0=0.7$. The strength of the magnetic field is denoted by the red triangles.
}
\label{magnetic}
\end{figure}

Let us set up a cylindrical coordinate system ($r,\theta,y$). Generally, the magnetic field of a point within the sphere at the $r$ ($r\leq R$) away from the center can be given by:
\begin{equation}
\vec{B}=\frac{2\rho_Q\vec{\omega}r^2}{15}\cos\theta\vec{e_r}+\frac{\rho_Q\vec{\omega}r^2}{15}\sin\theta\vec{e_{\theta}}+\frac{2\rho_Q\vec{\omega}(R^2-r^2)}{3}\vec{e_y}
\end{equation}
where $\vec{\omega}$ is the angular velocity, $\rho_Q$ is the charge density and $R$ is the radius. Since the global polarization is always considered to be along the $y$ direction, so we focus on the $e_y$ component of last term in the above equation. Considering the relationship between the vorticity $\vec{\Omega}$ and angular velocity $\vec{\omega}$:
\begin{equation}
\vec{\Omega}=\nabla\times\vec{\upsilon}=2\vec{\omega},
\end{equation}
then we will have:
\begin{equation}
\vec{B}_y=\frac{\rho_Q\vec{\Omega}(R^2-r^2)}{3}\vec{e_y}.
\end{equation}
Then according to ref. \cite{GLW2019}, at the freeze-out, one expects a polarization difference as:
\begin{equation}
\Delta P_m=\langle P_{\bar{\Lambda}}-P_{\Lambda}\rangle\simeq\langle\frac{2|\mu_{\Lambda}|\bar{\vec{B}}}{T_{fo}}\rangle,
\end{equation}
where $|\mu_{\Lambda}|=0.613\mu_N=\frac{0.613e}{2 M_N}$ with $M_N=938MeV$ and $T_{fo}$ is the freeze-out temperature (see Fig \ref{TempDen}). The $\langle\cdots\rangle$ is average over the space within the spherical collision system. The calculated polarization splitting $\Delta P_m$ as a function of the collisions energy is shown in Fig. \ref{magnetic}. As one can see from the Fig \ref{magnetic}, the polarization difference $\Delta P_m$ decreases with increasing collison energy and, is really trivial. Take the case of 7.7 GeV as an example: the polarization splitting is as low as $\Delta P_m=0.038\%$, far away from the experimental results. Comparing to  the results in Ref. \cite{GLW2019}, our results are one order of magnitude smaller and approaches the lower boundary of Ref. \cite{GLW2019}'s Fig. 3. We attribute it to the small magnitude of the induced magnetic field, as shown by the red triangles in our Fig. \ref{magnetic}. Actually due to the assumption of spherical collision system, instead of just a 'disk', the magnetic field calculated in our model is rather weak, which therefore results into a trivial results. Thus we can conclude that the magnetic field induced by the rotational charged current imposes a rather trivial effect on the $\Lambda$ and $\bar{\Lambda}$ polarization splitting, which should not be a surprise according to the prediction from magneto-hydrodynamics.

\subsection{Meson field}\label{S2.2}

According to Ref. \cite{CKW2019}, if the baryon current $\vec{J}_B=\rho_B\vec{\upsilon}$ is rotational, by assuming global equilibrium, one gets the magnetic field of meson:
\begin{equation}
\vec{B_V}=\frac{\bar{g}_V}{m_V^2}(\nabla\times \vec{J_B})= 
\frac{\bar{g}_V}{m_V^2}(\rho_B\vec{\Omega})
\end{equation}
where $\bar{g}_V$ is the vector meson's coupling constant, $\rho_B$ is the baryons density and $\vec{\Omega}=\nabla\times\vec{\upsilon}$ is the baryon vorticity.

If local equilibrium is assumed, the curl the baryon current would be :
\begin{equation}
\nabla\times\vec{J}=\rho_B\vec{\Omega}+\nabla\rho_B\times\vec{\upsilon}
\end{equation}
and then one gets the polarization splitting : 
\ba
\Delta \vec{P}_{J} = 
\langle C\frac{\vec{\nabla} \times \vec{J}_{\rm B}}{T} \rangle
&&= C\langle\frac{\rho_{\rm B} \, \vec{\Omega}}{T} \rangle
+ C\langle\frac{ \vec{\nabla}\rho_{\rm B} \times \vec{v} }{T} \rangle\, \nonumber \\
&&=\Delta\vec{P}_{\omega}+\Delta\vec{P}_{\rho}.
\label{mesondeltaP}
\ea
where $C=2(g_{\rm VH}\bar{g}_{\rm V})/(M_{\rm H} m_{\rm V}^2)$ is a coefficient determined
by strong coupling constants, hyperon and meson mass, and $\langle\cdots\rangle$ is the average over the space. The above coefficients are the same as in Ref. \cite{CKW2019}: $M_{\rm \Lambda}=1115.6MeV$, $M_{\rm V}=780MeV$, $\bar{g}_{\rm V}=5$, $g_{\rm V\Lambda}\approx 0.55g_{\rm VN}\approx 4.76$. From the above equation, the overall polarization splitting induced by the rotating baryon current, $\Delta \vec{P}_J$, is consist of two terms: the first term, $\Delta \vec{P}_{\omega}$ arises from the classical vorticity; the second term $\Delta \vec{P}_{\rho}$ is owed to the baryon density gradient.  

\begin{figure}
\begin{center}
\includegraphics[width=0.5\textwidth]{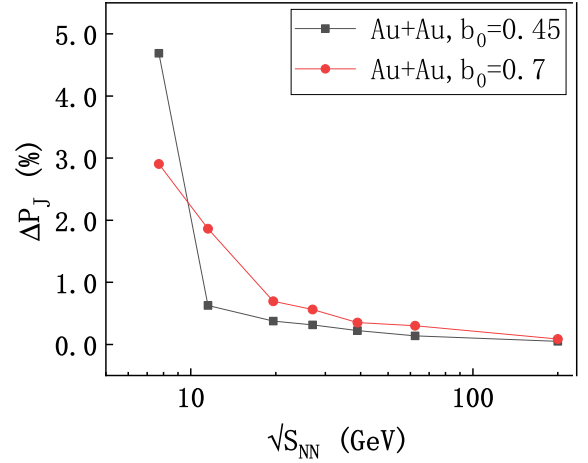}
\end{center}
\caption{
(Color online)
The overall polarization splitting $\Delta P_J$ induced by meson's magnetic field for different collision energies $\sqrt{S_{NN}}=7.7-200GeV$. The black line and red line correspond respectively the the cases of $b_0=0.45$ and $b_0=0.7$.}
\label{mesonj}
\end{figure}

In our previous work \cite{XCC2021}, the polarization splitting was calculated with impact parameter ratio $b_0=0.7$, which corresponds to the upper limit of centrality range $c = 20\%-50\%$ at STAR. In this work, the case with impact parameter ratio $b_0=0.45$, which corresponds to the lower limit of STAR centrality range, is additionally calculated. The results are shown in Fig. \ref{mesonj}, which shows the overall polarization splitting $\Delta \vec{P}_J$ induced by the meson's magnetic field. The black line denotes the polarization splitting at $b_0=0.45$ and the red line corresponds to the polarization splitting at $b_0 = 0.7$. Note here that due to the optimization of our initial state parameter, the results herein is a bit larger compared with our previous work \cite{XCC2021}, while all the characteristics and tendency are kept the same, including the behavior that the polarization splitting at 7.7 GeV decreases with the centrality. This behavior has been explained in our previous work in details\cite{XCC2021}, and we want to point out that similar behavior also occurs in UrQMD model \cite{VBZ2020}. For the 7.7GeV, a simple average of $\Delta P_J = 2.9\%$ ($b_0 = 0.45$) and $\Delta P_J = 4.6\%$ ($b_0 = 0.7$) gives a value of about $3.7\%$, which is significant and comparable with the experimental results. However, this value is still not as large as the experimental measured average value of about $6\%$. Therefore, as mentioned in the Introduction, the combination of different mechanisms to account might be necessary.

\section{Freeze-out space-time}\label{S3}

Just hinted by some polarization studies by AMPT model \cite{GSF2019} and UrQMD model \cite{VBZ2020}, the particles and  anti-particles in heavy ion collisions might be located in different collision region with different number/energy density, then naturally the freeze-out(FZ) time might be different. Since no theoretical study has been dedicated to this problem, we are not sure the magnitude of the FZ time difference. Therefore based on the Ref. \cite{VBZ2020}, a small but non-zero difference of freeze-out time between  $\bar{\Lambda}$ and $\Lambda$ in our PICR hydro model is assumed. More specifically, we assume that $\bar{\Lambda}$ particles freeze out earlier than $\Lambda$ particles, with two scenarios being considered: (1) the FZ time difference is $\Delta t = 0.423 *4 \approx 0.169$ fm/c, which is four time-steps of our PICR hydro model; (2) the FZ time difference is 16 time-steps $\Delta t=0.169 *4 \approx 0.677$ fm/c. 

\begin{figure}[ht] 
\begin{center}
      \includegraphics[width=8.5cm]{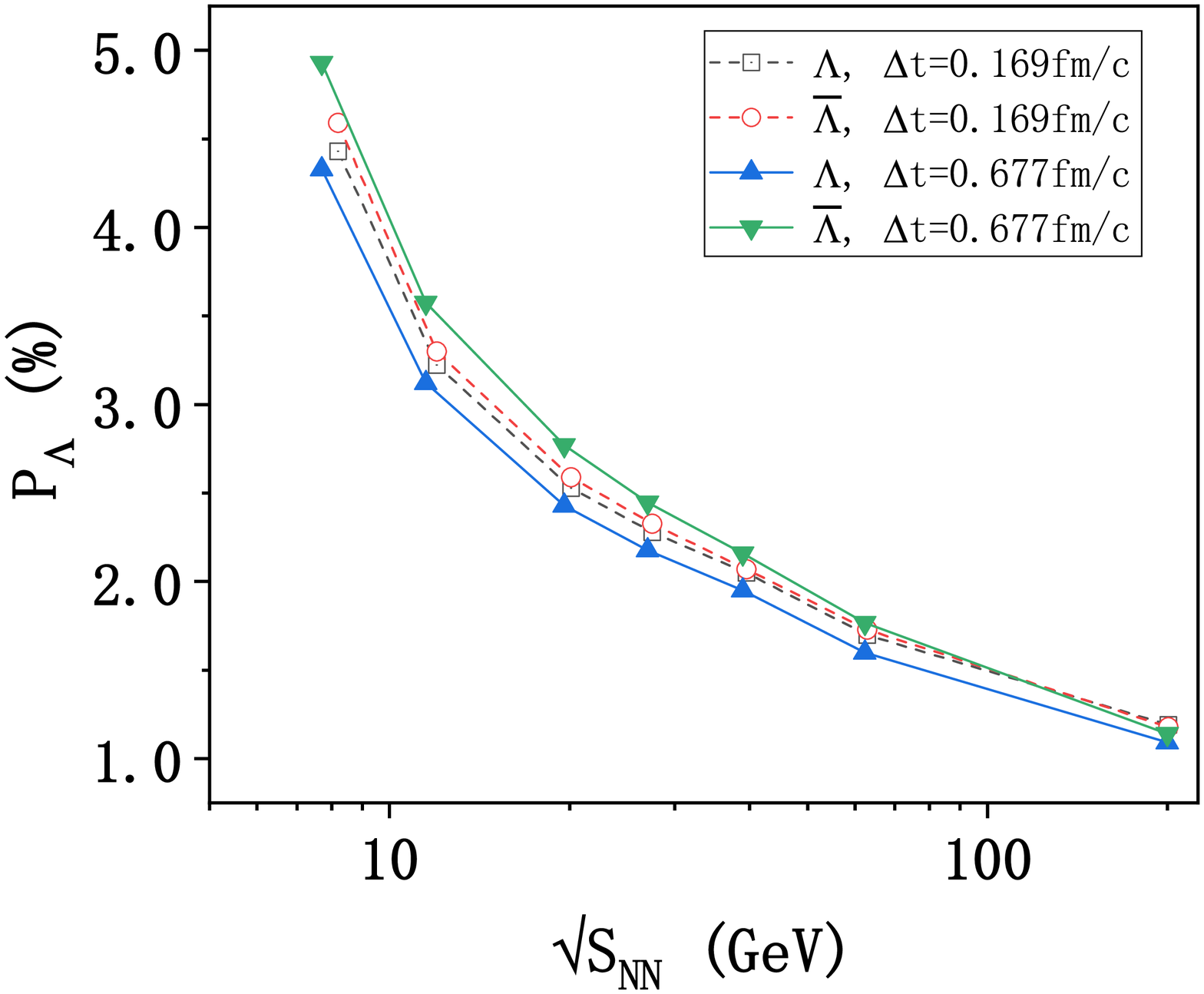}
      \includegraphics[width=8.5cm]{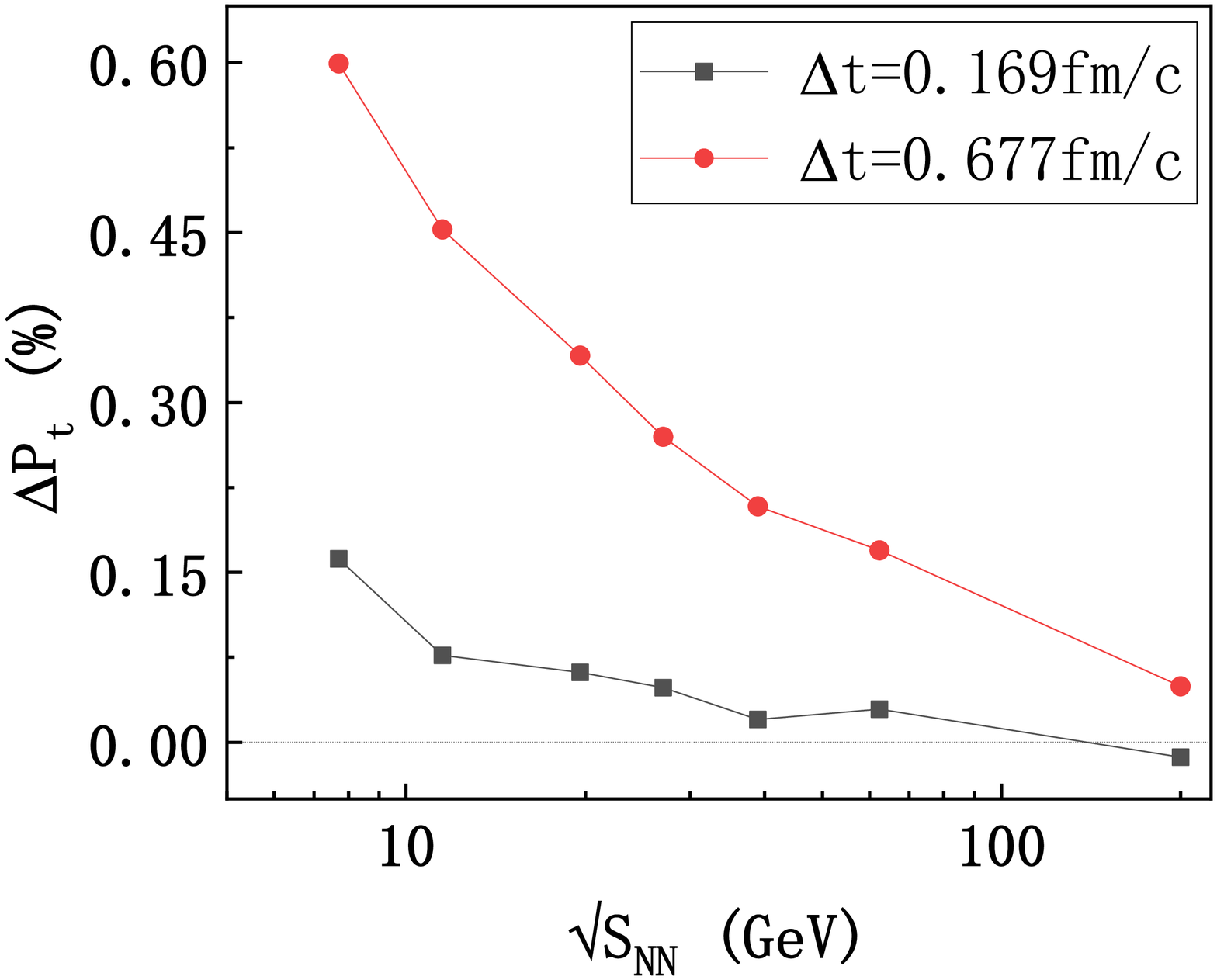}
\end{center}
\caption{
(Color online)
The upper panel shows the polarization of $\Lambda$ and $\bar{\Lambda}$ calculated in our PICR hydro model, based on the mechanism of freeze-out time difference.Two scenarios are shown here: the case of $\Delta t = 0.169$ fm/c is denoted by the hollow symbols; the case of $\Delta t = 0.667$ fm/c is denoted by the solid triangles. Note that the symbols for $\Delta t = 0.169 fm/c$ are all shifted to upper right a bit for purpose of distinction. The bottom panel shows directly the polarization difference $\Delta P_t=P_{\bar{\Lambda}} - P_{\Lambda}$, arises from FZ time difference.
}
\label{spacetime}
\end{figure}

Then we calculate the polarization of $\bar{\Lambda}$ and $\Lambda$ at fix impact parameter $b_0=0.7$, as shown in upper panel of Fig. \ref{spacetime}. The polarization difference induced by the FZ time difference, $\Delta P_t=P_{\bar{\Lambda}}-P_{\Lambda}$, has also been calculated and shown in the bottom panel of Fig. \ref{spacetime}. Comparing with the results in Ref. \cite{VBZ2020}, two things are worth to be noted: 1) the polarization difference in our model decreases with the increasing collision energy, while in Ref. \cite{VBZ2020} it is almost the same for different collision energy; 2) for two scenarios of small and large FZ time difference $\Delta t \approx 0.169$ fm/c  and $0.677$ fm/c , the polarization difference at 7.7 GeV approaches about $0.15\%$ and $0.6\%$, which is the same order of magnitude in Ref. \cite{VBZ2020} . By the way, if one expect the FZ time effect can solely and completely account the polarization splitting at 7.7 GeV, the FZ time difference needs to be about $2$ fm/c, which is unrealistic since the whole evolution time in our PICR hydro model is only 5.9 fm/c at 7.7 GeV.


\section{The total effect and Discussions}\label{S4}

\begin{figure}[ht]
\begin{center}
\includegraphics[width=8.5cm]{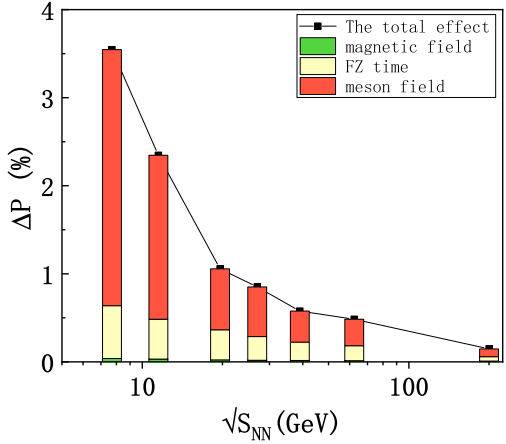}
\end{center}
\caption{
(Color online)
The total polarization splitting $\Delta P$ by putting together the three effects, including the $\Delta P_m$ denoted by green bottom columns, the $\Delta P_J$ denoted by the red top columns, and $\Delta P_t$ denoted by the middle yellow columns. We only show herein the case of larger FZ time difference $\Delta t\approx0.667$ fm/c, with the freeze-out time varying among $t_{fo}=5.9-7.9$ fm/c and the impact parameter ratio being fixed as $b_0=0.7$.}
\label{total2}
\end{figure}

Finally, we put the above three effects together for comparison,
$$\Delta P = \Delta P_m + \Delta P_J + \Delta P_t$$
where the correlated effect between the FZ time difference and the other two mechanisms is not considered, since we estimate it as only a small correction. The results are shown in Fig. \ref{total2}, where we only consider the case of $b_0 = 0.7$ and $\Delta t = 0.677$ fm/c.

The black line in Fig. \ref{total2} denotes the polarization splitting $\Delta P$ induced by total effect, red columnar part is polarization splitting $\Delta P_J$ induced by meson 'field', the yellow columnar part is polarization splitting $\Delta P_t$ induced by FZ time different and the green columnar part is polarization splitting $\Delta P_m$ induce by the QGP's magnetic field. We can see that the meson field effect is so dominant and accounts for the largest proportion. 

Taking the case of 7.7GeV for example: the meson field effect part contributes about 3\% to the total splitting effect, the FZ time effect is about 0.6\% and the magnetic field effect is only about 0.04\%,  so the hierarchy is $\Delta P_J \gg \Delta P_t \gg \Delta P_m$, as shown in Fig. \ref{total2}. We can see that the polarization splitting induced by the magnetic field effect part is rather trivial and the dominant effect that induces polarization splitting is the meson field. The main reason is simple: the meson (magnetic) field is driven by strong coupling interaction, which is an order of magnitude larger than the electromagnetic interaction. On the other hand, the contribution from FZ time effect is just right between the QGP's magnetic field and the meson magnetic field effect, because it is limited by the assumption that the difference of FZ time between $\Lambda$ and $\bar{\Lambda}$ should be small. 

\begin{figure}[htbp]
\begin{center}
\includegraphics[width=8.5cm]{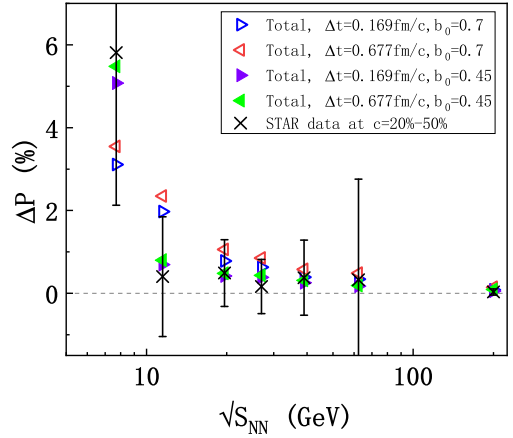}
\end{center}
\caption{
(Color online)
The total polarization splitting between $\Lambda$ and $\bar{\Lambda}$ for different collisions energies $\sqrt{S_{NN}}=7.7-200$ GeV, with different impact parameter ratio $b_0=0.45$ (solid triangles) $b_0=0.7$ (hollow triangles), different FZ time difference $\Delta t = 0.169$ fm/c (left-directed triangles) and $\Delta t = 0.677$ fm/c (right-directed triangles). For comparison the experimental data are also shown, by the cross symbols with error bars \cite{SecondSTAR,STAR2018,STAR2021}. (Feed-down effect was considered \cite{XXH2019,BCS2019}, and decaying parameter was renewed as $\alpha = 0.732$). 
}
\label{experiment}
\end{figure}

Finally, we compare our results with the experimental data from STAR collaboration \cite{SecondSTAR,STAR2018,STAR2021} as shown in Fig. \ref{experiment}. We can see that our results still agrees fairly well with the experimental data denoted by cross symbols. Specifically for the 7.7 GeV, comparing to our previous work where the polarization splitting induced by meson field was only about 3\%-4\%, while now combining the other two effects, the splitting could hike to 5.5\% at most, with a mean value of about 4.5\%. In another words, the combined effect could enhance the polarization difference by about 10\%-20\%, to an average value of 4\%-4.5\% polarization splitting, and this is more close to the mean value of 5.8\% measured by STAR. We expect that with more mechanisms being combined on the basis of a uniform model, the splitting effect could be accounted satisfactorily.

\section{Summary and Conclusion}\label{S5}
The STAR's measurement on the $\Lambda$ polarization at 7.7GeV collisions shows a significant splitting of $\Lambda$ and $\bar{\Lambda}$ polarization at Au+Au 7.7GeV collisions. This huge signal seems unable to be described satisfactorily by any single mechanism by far, thus we revisit and then combine there different mechanisms together on the basis of our PICR hydrodynamic model, to explain the experimental data. The three mechanisms, i.e. the meson field mechanism, the freeze-out space-time mechanism, and the QGP's magnetic field mechanism, lie on different stage of high energy collisions. 

By revisiting the 3 mechanisms, we find a hierarchy of $\Delta P_J \gg \Delta P_t \gg \Delta P_m$, and: 

(1) the QGP's magnetic field mechanism is rather trivial. Our calculation based on the the assumption of sphere, instead of a 'disk', signals a very weak magnetic field, which actually corresponds to the lower boundary of magnetic field in Ref. \cite{GLW2019}. Therefore, it results into a trivial splitting effect.

(2) the splitting arises from freeze-out time effect is larger than that from the fluid magnetic field mechanism, by an order of magnitude, while it is also smaller than the meson field mechanism by an order of magnitude, since it is restricted by the small FZ time difference.

(3) the meson field mechanism is dominant. Besides, we reconfirm the interesting behavior that collisions of lower centrality ($b_0 = 0.45$) have a larger polarization splitting than that of higher centrality ($b_0 = 0.7$). This behavior actually can be also seen in the UrQMD model of Ref. \cite{VBZ2020} (see Fig. 8 therein).

Besides, by combing the 3 mechanisms, we could promote the mean value of polarization splitting from 3\%-4\% to 4.5\%, which is more close to the experimental measured mean value of about 5.8\%.

\section*{Acknowledgments}

The work of of Y. L. Xie is supported by National Natural Science Foundation of China (No. 12005196).

\end{document}